\documentclass[
reprint,
superscriptaddress,
nobibnotes,
 amsmath,amssymb,
 aps,
 prl,
 floatfix,
]{revtex4-2}

\usepackage{graphicx}
\usepackage{epstopdf} 
\usepackage{dcolumn}
\usepackage{bm}
\usepackage{siunitx}
\usepackage{hyperref}
\usepackage{xcolor}
\usepackage{float}

\begin{document}

\title{Ferromagnetic resonance and magnetic anisotropy in YbMn$_\text{6}$Sn$_\text{6}$}
\author{Philipp Schwenke}
\email[E-Mail: ]{philipp.schwenke@rptu.de}
\affiliation{Fachbereich Physik and Landesforschungszentrum OPTIMAS, Rheinland-Pf{\"a}lzische Technische Universit{\"a}t Kaiserslautern-Landau, 67663 Kaiserslautern, Germany}

\author{Kyle W. Fruhling}
\affiliation{Department of Physics, Boston College, Chestnut Hill, Massachusetts 02467, USA}

\author{David Weffling}
\affiliation{Fachbereich Physik and Landesforschungszentrum OPTIMAS, Rheinland-Pf{\"a}lzische Technische Universit{\"a}t Kaiserslautern-Landau, 67663 Kaiserslautern, Germany}

\author{Vitaliy I. Vasyuchka}
\affiliation{Fachbereich Physik and Landesforschungszentrum OPTIMAS, Rheinland-Pf{\"a}lzische Technische Universit{\"a}t Kaiserslautern-Landau, 67663 Kaiserslautern, Germany}

\author{Fazel Tafti}
\affiliation{Department of Physics, Boston College, Chestnut Hill, Massachusetts 02467, USA}

\author{Mathias Weiler}
\affiliation{Fachbereich Physik and Landesforschungszentrum OPTIMAS, Rheinland-Pf{\"a}lzische Technische Universit{\"a}t Kaiserslautern-Landau, 67663 Kaiserslautern, Germany}

\begin{abstract}

The RMn$_\text{6}$Sn$_\text{6}$ (R = rare earth) family of kagome magnets exhibits a rich variety of magnetic states and unconventional electronic and magnetic properties. While the static magnetic properties of these materials have been extensively investigated, their magnetization dynamics remain less explored. Here, we investigate the magnetic anisotropy and dynamic magnetization of ferromagnetic YbMn$_\text{6}$Sn$_\text{6}$ using angle-dependent ferromagnetic resonance (FMR) spectroscopy and DC magnetization measurements. The angular dependence of FMR reveals the expected uniaxial anisotropy along the crystallographic c-axis, which is the global magnetically hard axis. In addition, we identify a pronounced twofold anisotropy within the ab-plane, which is independently confirmed by angle-dependent DC magnetization measurements. The observed uniaxial anisotropy in the ab-plane indicates a breaking of the expected sixfold rotational symmetry in the magnetic response that is attributed to growth induced anisotropy.

\end{abstract}

\maketitle

Kagome magnets have attracted considerable interest due to the interplay of geometric frustration, electronic correlations, and nontrivial band topology \cite{bolens_topological_2019, chisnell_topological_2015, zhang_topological_2020, zhang_topological_2023, yin_topological_2022, yin_negative_2019, wang_quantum_2023}.
The RMn$_\mathrm{6}$Sn$_\mathrm{6}$ (R = rare earth) family is a prominent example, consisting of layered structures with Mn kagome planes separated by rare-earth-containing triangular layers. 
Depending on the choice of the rare-earth element, these materials exhibit a variety of magnetic ground states, including ferromagnetic, collinear antiferromagnetic, and incommensurate spin-spiral order \cite{lv_anomalous_2023, l_high-field_2025, li_discovery_2023, mozaffari_diverse_2025, samatham_perturbation-tuned_2024, clatterbuck_magnetic_1999, venturini_incommensurate_1996, zhang_magnetic_2022, ghimire_competing_2020, dally_chiral_2021, fruhling_characterization_2024}.
Their combination of magnetic order and nontrivial electronic structure gives rise to a range of unconventional transport and magnetic phenomena, including anomalous and topological Hall effects as well as topological magnetic textures \cite{mozaffari_diverse_2025, jiang_anomalous_2024, li__enhanced_2024, emmanuel_ermn6sn6_2025, ma_large_2026, jones_origin_2024, madhogaria_topological_2023, zhang_magnetic_2022, xu_topological_2022, yin_quantum-limit_2020, li_discovery_2023, fruhling_topological_2024}

As a member of the RMn$_\mathrm{6}$Sn$_\mathrm{6}$ material family, YbMn$_\mathrm{6}$Sn$_\mathrm{6}$ crystallizes in a layered structure consisting of Mn kagome layers alternating with Yb/Sn layers.
This compound exhibits a ferromagnetic order below its Curie temperature at around $T_\mathrm{C} = \SI{300}{\kelvin}$ \cite{jiang_anomalous_2024, li__enhanced_2024, mazet_study_1999}.
The magnetic order is attributed to the Mn sublattice, while no magnetic order has been observed on the Yb sublattice \cite{eichenberger_commensurate-incommensurate_2017, magnette_crystal_2018}.
Previous studies have established a pronounced anisotropy between the crystallographic c axis and the ab plane, with the c-axis reported as the magnetic hard axis and the ab-plane as the easy plane \cite{jiang_anomalous_2024, li__enhanced_2024, lv_anomalous_2023}.
However, the magnetic anisotropy within the ab plane has received considerably less attention.
\begin{figure}[h]
\centering
\includegraphics{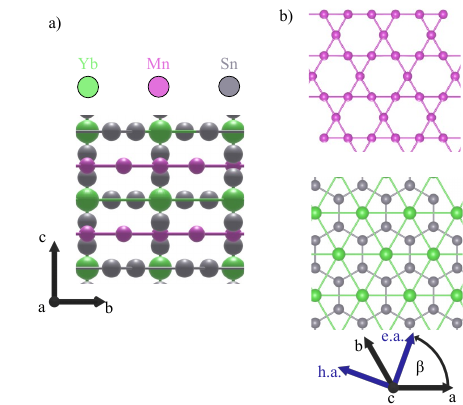}
\caption{
Side view of the crystal structure of YbMn$_\text{6}$Sn$_\text{6}$ a). The green spheres represent the Yb atoms, the purple ones the Mn atoms and the gray spheres the Sn atoms.
In b) the top view of the sublattices of the crystal are illustrated. 
The top sublattice corresponds to the Kagome Mn lattice and the bottom to the triangular Yb and Sn sublattice.
The black coordinate system in the bottom indicates the a, b, and c-axes.
The blue arrows together with the c-axis describe the coordinate system used within this work.
There, e.a. indicates the magnetic easy axis and h.a. the magnetic hard axis in the ab-plane.
This coordinate system is at an unknown angle $\beta$ to the a axis.
The images are created using the software vesta \cite{momma_vesta_2011} and using the atomic coordinates from Ref.~\cite{xia_ybmn6_2006}.
}
\label{fig:YbMn6Sn6}
\end{figure}
Angle-dependent ferromagnetic resonance spectroscopy (FMR) provides a sensitive probe of magnetic anisotropy, as the resonance condition depends strongly on the magnetic free-energy landscape. 
Recent FMR measurements on MgMn$_\mathrm{6}$Sn$_\mathrm{6}$ have demonstrated the suitability of FMR for investigating the dynamic response and magnetic anisotropy in this material family, establishing the uniaxial anisotropy between the c-axis and the ab-plane \cite{pal_ferromagnetic_2025}.
In this work, we investigate the magnetic anisotropy of YbMn$_\mathrm{6}$Sn$_\mathrm{6}$ using angle-dependent FMR and DC magnetometry. While the measurements reveal the expected uniaxial anisotropy distinguishing the c axis from the ab plane, we additionally observe a pronounced twofold anisotropy within the ab plane.
The observed twofold ab-plane anisotropy is not consistent with the expected sixfold rotational symmetry of the hexagonal crystal structure and demonstrates that the magnetic response of YbMn$_\mathrm{6}$Sn$_\mathrm{6}$ is not fully captured by a simple c-axis uniaxial anisotropy and shape anisotropy.

\begin{figure*}[ht]
\centering
\includegraphics{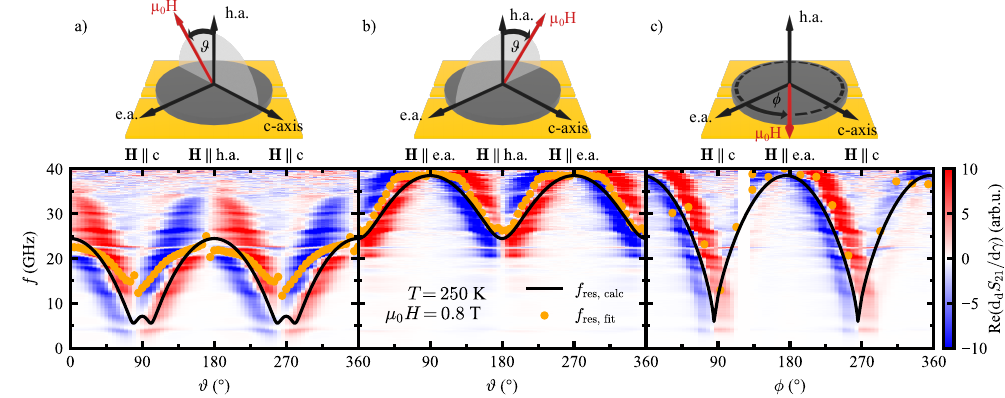}
\caption{
Frequency $f$ and angle dependence of the transmission parameter $S_{21}$ after performing background subtraction via derivative divide \cite{maier-flaig_note_2018} (with  $\gamma = \vartheta,\phi$) at $\mu_\text{0}H = \SI{0.8}{\tesla}$.
The orange dots represent the fitted resonance position $f_\text{res, fit}$ while the black line depicts the calculated resonance position $f_\text{res, calc}$ by Eq.~\eqref{eq:Smit}.
The coordinate system is given by the easy-, hard- and c-axis of the system.
In a) the magnetic field is rotated in the plane set by the c-axis and hard axis in the ab-plane.
The rotation plane corresponding to the ab-plane of the crystal, spanned by the easy- and hard axis is depicted in b), and in c) the magnetic field rotation within the easy-axis and c-axis plane is shown.
}
\label{fig:AdFMR}
\end{figure*}

The crystal structure of YbMn$_\mathrm{6}$Sn$_\mathrm{6}$ is shown in Fig.~\ref{fig:YbMn6Sn6}. 
The structure was visualized using the software VESTA~\cite{momma_vesta_2011} based on the atomic coordinates reported in Ref.~\cite{xia_ybmn6_2006}. 
Figure~\ref{fig:YbMn6Sn6}~a) presents a side view along the crystallographic $a$-axis, highlighting the layered nature of the crystal. 
The Mn kagome layers and the Yb-containing layers are stacked alternately along the $c$-axis. 
It should be noted that the Yb layer and the Sn sites in between the Mn- and the Yb-containing layers are partially occupied \cite{xia_ybmn6_2006} which is not represented in the Figure.
Figure~\ref{fig:YbMn6Sn6}~b) shows the corresponding top view along the $c$-axis. 
The upper panel displays the Mn kagome lattice, while the lower panel illustrates the triangular arrangement of the Yb atoms.
The coordinate system depicts the crystallographic axes in black and the coordinate system spanned by the magnetic easy and hard axis in blue. 
The angle $\beta$ between these coordinate systems is unknown.
To describe the magnetization dynamics of the material, we start from the magnetic free energy $F$ given by
\begin{equation}
    F = F_\text{Zee}+F_\text{demag}+F_\text{ani}
    \label{eq:F}
\end{equation}

where $F_\text{Zee}$ denotes the Zeeman energy, $F_\text{demag}$ the demagnetization (shape anisotropy) energy, and $F_\text{ani}$ the magnetocrystalline anisotropy energy.

The ferromagnetic resonance frequency can then be calculated from the free energy using \cite{smit_ferromagnetic_1955, suhl_ferromagnetic_1955}

\begin{equation}
    f = \frac{\gamma}{2\pi M_\text{s}\sin(\vartheta_\text{0})}\sqrt{F_{\vartheta\vartheta}F_{\phi\phi}-F_{\vartheta\phi}F_{\phi\vartheta}},
    \label{eq:Smit}
\end{equation}

where $F_{ij}=\partial^2F/\partial i,\partial j$ are the second derivatives of the free energy evaluated at the equilibrium orientation of the magnetization, which is determined by the condition $\partial F/\partial\phi = \partial F/\partial\vartheta = 0$ from minimization of the free energy.
It should be noted that Eq.~\eqref{eq:Smit} is only defined for $\vartheta_\text{0}\neq0$.
Consequently, the coordinate system must be chosen such that $\vartheta_\text{0}\neq0$ for all magnetization orientations as described in detail in the Supplementary Material \cite{supplement}.

To determine the crystallographic orientations of the sample we performed angle-dependent FMR (AdFMR) measurements.
Figure~\ref{fig:AdFMR} shows a schematic of the magnetic field rotation (top) together with the corresponding frequency- and angle-dependent, background-corrected transmission parameter $S_{21}$ (bottom). 
The measurements were carried out at a temperature of $T=\SI{250}{\kelvin}$, which is sufficiently below the Curie temperature to ensure that the sample is well within the ferromagnetically ordered phase. 
The external magnetic field was fixed at $\mu_0H=\SI{0.8}{\tesla}$, as higher magnetic fields would shift the resonance frequency beyond the accessible measurement range.
The orange dots indicate the fitted resonance frequencies, $f_\mathrm{res,fit}$, while the black line represents the resonance frequencies, $f_\mathrm{res,calc}$, calculated using Eq.~\eqref{eq:Smit}.
The parameters used for the calculation are $M_\mathrm{s} = \SI{350}{\kilo\ampere\per\meter}$, $g = 2.05$, $K_\mathrm{u} = \SI{150}{\kilo\joule\per\meter\cubed}$, $K_\mathrm{u2} = \SI{15}{\kilo\joule\per\meter\cubed}$, $N_{xx} =0.15$, $N_{yy} = 0.21$ and $N_{zz} = 0.62$.
Where $K_\mathrm{u}$ is the magnetocrystalline anisotropy constant along the x-axis and $K_\mathrm{u2}$ the anisotropy constant for the two-fold symmteric anisotropy within the ab-plane.
The parameters for $g$, $K_\mathrm{u}$ and $K_\mathrm{u2}$ are manually varied until good agreement between calculation and experiment is reached.
$M_\mathrm{s}$ is extracted from magnetization data from a crystal of the same fabrication batch and $N_{jj}$ with $j = x,y,z$ are determined by approximating the sample as an ellipsoid and calculating the demagnetization factors.

To calculate $f_\mathrm{res,calc}$ for the out-of-plane rotation, the coordinate system is rotated in the calculation to ensure that the condition $\vartheta_0\neq\SI{0}{\degree}$ is fulfilled. 
Furthermore, the equilibrium orientation of $M$ is obtained by numerically minimizing the free energy $F$.
In Figure~\ref{fig:AdFMR}~a), the magnetic field is rotated within the plane spanned by the crystallographic $c$-axis and the hard axis of the ab-plane (h.a.).

The angular dependence of the resonance frequency exhibits a $\SI{180}{\degree}$ periodicity, with minima at $\vartheta=\SI{90}{\degree}$ and $\vartheta=\SI{270}{\degree}$. 
This behavior indicates the presence of a magnetic hard axis, which we attribute to the crystallographic $c$-axis in agreement with previous reports \cite{lv_anomalous_2023, jiang_anomalous_2024}.
Overall, the calculated resonance frequencies reproduce the experimental data well. 
However, a noticeable deviation between the calculated resonance frequency, $f_\mathrm{res,calc}$, and the fitted resonance frequency, $f_\mathrm{res,fit}$, is observed when the magnetic field is applied parallel to the $c$-axis. 

Since the actual sample geometry deviates from an ellipsoid used in the calculation, the corresponding demagnetization energy is expected to differ slightly. 
As a consequence, the equilibrium orientation of the magnetization may deviate from the calculated one, resulting in the observed discrepancy in the resonance frequency.

For $\vartheta=\SI{0}{\degree}$, good agreement between $f_\mathrm{res,fit}$ and $f_\mathrm{res,calc}$ is obtained. 

Figure~\ref{fig:AdFMR}~b) shows the AdFMR measurement for a magnetic field rotation in a plane perpendicular to the previously identified crystallographic $c$-axis. 
The rotation starts at $\vartheta=\SI{0}{\degree}$, corresponding to the same field orientation as in Figure~\ref{fig:AdFMR}~a). 
As the magnetic field is rotated towards $\vartheta=\SI{90}{\degree}$, the resonance frequency increases, indicating that the field is aligned with a magnetic easy axis ($\mathbf{H}\parallel\mathrm{e.a.}$). 

Over the entire angular range, the calculated resonance frequencies, $f_\mathrm{res,calc}$, are in good agreement with the fitted resonance frequencies, $f_\mathrm{res,fit}$.
Figure~\ref{fig:AdFMR}~c) presents the AdFMR data for magnetic field rotation within the plane of the crystal (lower panel), together with a schematic of the rotation geometry (upper panel). 
The rotation starts with the magnetic field aligned along the easy axis ($\mathbf{H}\parallel\mathrm{e.a.}$). 
Upon rotating the field by $\phi=\SI{90}{\degree}$, the field becomes parallel to the crystallographic $c$-axis ($\mathbf{H}\parallel c$). 
The white region around $\phi\approx\SI{125}{\degree}$ originates from missing data points caused by the background subtraction procedure and the subsequent correction of the measurement angles to the coordinate system shown in the upper panel of Figure~\ref{fig:AdFMR}~c). 
In addition, the reduced signal intensity for $\phi=\SI{90}{\degree}$--$\SI{170}{\degree}$ and $\phi=\SI{270}{\degree}$--$\SI{360}{\degree}$ results from the reduced excitation efficiency of the coplanar waveguide (CPW), as the in-plane component of the microwave driving field becomes nearly parallel to the external magnetic field.
The resonance frequency reaches its maximum for $\phi=\SI{0}{\degree}$ and its minimum for $\phi=\SI{90}{\degree}$, consistent with the magnetic field being rotated from the easy axis to the hard $c$-axis. 
The calculated resonance frequencies reproduce the experimental data well over the entire angular range, demonstrating that the free-energy model provides an accurate description of the magnetic anisotropy. 

From these AdFMR measurement data a magnetic easy-axis (e.a.), a magnetic hard axis (c-axis) and an intermediate hard axis (h.a.) that is located in the ab-plane are distinguishable.
While the uniaxial anisotropy along the c-axis is known \cite{lv_anomalous_2023, jiang_anomalous_2024, li__enhanced_2024},
the additional uniaxial anisotropy within the ab-plane is surprising considering the threefold symmetry of the crystal lattice.
Although shape anisotropy provides the dominant contribution of the uniaxial anisotropy in the ab-plane, a small cristallographic anisotropy is required to fully reproduce the AdFMR data.
This additional anisotropy might be caused by strain in the crystal or a growth induced anisotropy but further investigations are necessary to determine the microscopic origin.

\begin{figure}[h]
\centering
\includegraphics{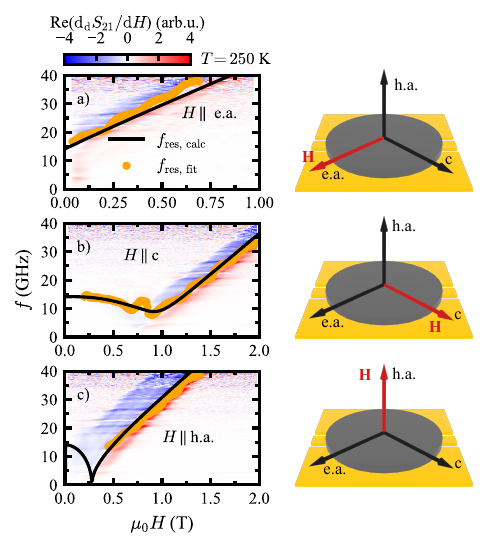}
\caption{
Magnetic field $\mu_\text{0}H$ and frequency $f$ dependence of the real part of the transmission parameter $S_{21}$ after background subtraction.
The magnetic field is applied along the easy-axis a), the c-axis b) and the hard-axis in the ab-plane c).
$\mu_\text{0}H$ is reduced from the maximum field to 0 in the measurements.
The orange dots represent the fitted resonance position $f_\text{res, fit}$ while the black line depicts the calculated resonance position $f_\text{res, calc}$ by Eq.~\eqref{eq:Smit}.
On the right schematics depicting the geometry of the applied magnetic field are shown.
}
\label{fig:FMR-axes}
\end{figure}

To validate the anisotropy constants obtained from the AdFMR measurements, field-dependent FMR measurements were performed with the external magnetic field aligned along the in-plane easy axis (e.a.), in-plane hard axis (h.a.), and the $c$-axis. 
The resulting background-corrected FMR spectra are shown in Figure~\ref{fig:FMR-axes} (left), together with the fitted resonance frequencies, $f_\mathrm{res,fit}$ (orange dots), and the calculated resonance frequencies, $f_\mathrm{res,calc}$ (black line). 
Schematics of the respective measurement geometries are shown on the right.

Figure~\ref{fig:FMR-axes}~a) illustrates the frequency and field dependence of the real part of the transmission parameter $S_{21}$ for $\mathbf{H}\parallel\mathrm{e.a.}$ after subtraction of the field-independent background.
During the measurement, $\mu_\mathrm{0}H$ is varied from the maximum applied field to zero to ensure that the sample is initially saturated along the field direction. 
$f_\mathrm{res, calc}$ are in good agreement with the experimentally determined resonance positions over the entire field range. 
In addition, the frequency linewidth $\Delta f$ was extracted from the spectra, yielding a Gilbert damping parameter of $\alpha=(0.113\pm0.002)$.
The frequency dependence of $\Delta f$ and the corresponding fit are presented in the Supplementary Material \cite{supplement}.

The corresponding measurements for $\mathbf{H}\parallel c$ are shown in Figure~\ref{fig:FMR-axes}~b).
In contrast to the easy-axis configuration, the resonance frequency exhibits a pronounced nonlinear field dependence below approximately $\mu_0H=\SI{1}{\tesla}$, reflecting the strong uniaxial anisotropy along the $c$-axis. 
Above this field, the resonance approaches the high-field regime. 
Once again, the calculated resonance frequencies accurately reproduce the experimental data.

Figure~\ref{fig:FMR-axes}~c) presents the field-dependent FMR spectra for the magnetic field applied along the in-plane hard axis ($\mathbf{H}\parallel\mathrm{h.a.}$). 
Also in this configuration, good agreement is obtained between $f_\mathrm{res,calc}$ and $f_\mathrm{res,fit}$.
It should be noted that the data indicate that an anisotropy field of approximately $\mu_0H_\text{ani} \approx\SI{0.28}{\tesla}$ is necessary to accurately model the data.
In contrast, the anisotropy field arising from the shape anisotropy alone is only $\mu_0H_\text{shape} = (0.206\pm0.012)\si{\tesla}$. This discrepancy suggests that an additional anisotropy contribution is required.
The excellent agreement between experiment and calculation for all three field orientations demonstrates that the extracted anisotropy constants provide a consistent description of the FMR response of YbMn$_\text{6}$Sn$_\text{6}$.

\begin{figure}[h]
\centering
\includegraphics{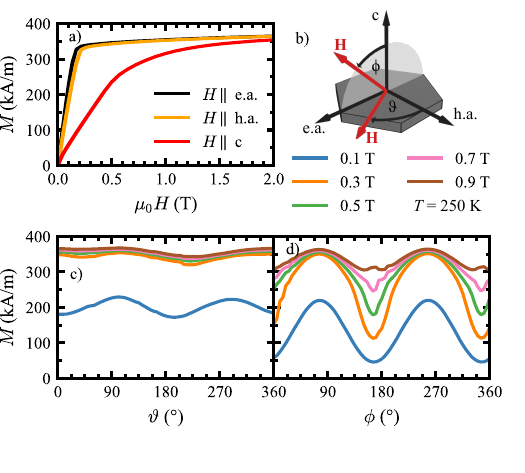}
\caption{
Magnetic field and angle dependence of the magnetization in a second YbMn$_\text{6}$Sn$_\text{6}$ crystal.
a) depicts the $\mu_\text{0}H$ dependence of $M$ with the magnetic field applied along the easy- (e.a. - black line), the hard- (h.a. - orange line) and along the c-axis (red line) at $T = \SI{250}{\kelvin}$.
In b) a schematic of the coordinate system and the rotation angles for the panels c) and d) are shown.
Additionally, the legend for the different colored lines in c) and d) is illustrated indicating the applied magnetic field $\mu_\text{0}H$ for each line.
The angle dependence of $M$ when rotating $\mathbf{H}$ in the ab-plane (e.a.-h.a. plane) is shown in c), while the angle dependence in the e.a.-c plane rotation is depicted in d).
These are illustrated for magnetic fields from $\mu_\text{0}H = \SI{0.1}{\tesla} -\SI{0.9}{\tesla}$ in $\Delta\mu_\text{0}H = \SI{0.2}{\tesla}$ steps.
}
\label{fig:VSM}
\end{figure}

To further support the anisotropy parameters obtained from the FMR measurements, DC magnetization measurements were performed on a second crystal from the same growth batch. 
Figure~\ref{fig:VSM}~a) shows the field-dependent magnetization, $M(H)$, for magnetic fields applied along the magnetic easy axis (black), the ab-plane hard axis (orange), and the crystallographic $c$-axis (red). 
Saturation is reached only at comparatively high magnetic fields for $\mathbf{H}\parallel c$, reflecting the strong uniaxial anisotropy along this direction. 
In contrast, the easy and ab-plane hard axes exhibit similar saturation fields, with the latter requiring a slightly larger field to reach saturation.
To obtain a more detailed picture of the magnetic anisotropy, angle-dependent DC magnetization measurements were carried out. 
Figure~\ref{fig:VSM}~b) shows a schematic of the crystal with the coordinate system used. 
The legend below the schematic indicates the applied magnetic field for each angular-dependent measurement.
The angular dependence of the magnetization, $M(\vartheta)$, measured at different magnetic fields is presented in ure.~\ref{fig:VSM}~c). 
At low magnetic fields, the magnetization exhibits a pronounced $\SI{180}{\degree}$ periodicity. 
This behavior contrasts with the sixfold rotational symmetry expected from the crystal structure and is instead consistent with the additional two-fold symmetric anisotropy in the ab-plane identified by the FMR measurements (compare Figure~\ref{fig:AdFMR}). 
At higher magnetic fields, the angular dependence becomes significantly weaker as the sample approaches magnetic saturation (see Figure~\ref{fig:VSM}~a)). 
Extracting the anisotropy constant yields $K_\mathrm{u,2}=\SI{7.2}{\kilo\joule\per\meter\cubed}$ for this crystal.
Furthermore, this angular dependence can not be explained by the shape anisotropy as the demagnetization factors for the second crystal are $N_{xx} =0.42$, $N_{yy} = 0.41$ and $N_{zz} = 0.16$ when assuming the sample as ellipsoid with the crystallographic c-axis along the $z$-direction.
Since $N_{xx}\approx N_{yy}$, the ab-plane is considered isotropic. 
Although the value of $K_\mathrm{u,2}$ is approximately a factor of two to three smaller than that obtained from the AdFMR measurements, the discrepancy is attributed to the measurements being performed on a different crystal from the same growth batch. 
Small variations in crystal properties or temperature dependence may therefore result in slightly different anisotropy constants. 
In addition, the FMR crystal has its crystallographic $c$-axis oriented within the sample plane, making the extracted anisotropy more sensitive to the assumed demagnetization correction.
It should be noted that the effect of the two-fold anisotropy in the ab-plane is significantly more pronounced in the FMR measurement (compare Figure \ref{fig:AdFMR}~b)) as the shape anisotropy has a significant contribution in the crystal used in the FMR measurements.

Figure~\ref{fig:VSM}~d) shows the angular dependence $M(\phi)$ for magnetic field rotation in the $e.a.-c$ plane. 
As expected, a $\SI{180}{\degree}$ periodicity is observed, reflecting the strong uniaxial anisotropy associated with the crystallographic $c$-axis.
Extracting the effective anisotropy constant yields $K_\mathrm{eff,1}=\SI{142.3}{\kilo\joule\per\meter\cubed}$. 
The contribution from the shape anisotropy is $K_\mathrm{shape} = \SI{-19.2}{\kilo\joule\per\meter\cubed}$.
Therefore, the resulting uniaxial anisotropy constant is given by $K_\mathrm{u,1} = K_\mathrm{eff}-K_\mathrm{shape} = \SI{161.5}{\kilo\joule\per\meter\cubed}$.
The extracted anisotropy constant, $K_\mathrm{u,1}$, is in good agreement with the value obtained from the FMR measurements on the other crystal.
Furthermore it is comparable to the value reported in MgMn$_\text{6}$Sn$_\text{6}$ at $T = \SI{300}{\kelvin}$ in Ref.~\cite{pal_ferromagnetic_2025}.
To verify that the observed ab-plane anisotropy is reproducible, equivalent angle-dependent DC magnetization measurements were performed on a third crystal from a different growth batch. The same twofold angular dependence is observed in the ab-plane of the crystal (see Supplementary Material \cite{supplement}).

In conclusion, we investigated the magnetization dynamics in a member of the RMn$_\text{6}$Sn$_\text{6}$ material family using ferromagnetic resonance spectroscopy and extracted the anisotropy along the c-axis an within the ab-plane. 
The pronounced uniaxial anisotropy associated with the crystallographic $c$-axis enables an unambiguous determination of the crystal orientation. 
In addition, we identify a previously unreported two-fold anisotropy in the ab-plane, which is independently confirmed by angle-dependent DC magnetization measurements on a second crystal from the same growth batch.
The observed twofold symmetry of the in-plane anisotropy is unexpected, as the hexagonal crystal symmetry would suggest a sixfold in-plane anisotropy, while previous studies have described the $ab$ plane as magnetically easy \cite{lv_anomalous_2023, jiang_anomalous_2024, li__enhanced_2024}. 
Our results therefore indicate that the magnetic anisotropy of YbMn$_\text{6}$Sn$_\text{6}$ is more complex than previously assumed and motivate further investigations into its microscopic origin.

\begin{acknowledgments}
\section{Acknowldegements}
We acknowledge financial support by the Deutsche Forschungsgemeinschaft (DFG, German Research Foundation) within the Transregional Collaborative Research Center TRR 173/3–268565370 ”Spin +X” (Project B13).
We acknowledge financial support from the European Research Council (ERC) under the European Union’s Horizon Europe research and innovation programme (Grant agreement No. 101044526).
The work at Boston College (crystal growth and DC magnetization measurements) was supported by the U.S. Department of Energy, Office of Basic Energy Sciences, Division of Physical Behavior of Materials under award number DE-SC0023124.
K.F. would like to thank Siddharth Nandandwar for assistance with crystal sample shape measurements.

\textit{Data availability}: The data that support the findings of
 this article are openly available \cite{schwenke_ybmn6sn6_2026}, embargo periods may
 apply.
\end{acknowledgments}

\bibliography{Bib}


\end{document}